\documentclass[useAMS,usenatbib,usegraphicx]{mn2e}

\usepackage{amsmath}

\title[]{A TWO ZONE SYNCHROTRON MODEL FOR THE KNOTS IN THE M87 JET}
\author[S. Sahayanathan]{S. Sahayanathan\thanks{E-mail:
sunder@barc.gov.in}\\
Astrophysical Sciences Division, Bhabha Atomic Research Centre, Mumbai - 400085, India\\}
\begin{document}



\maketitle

\label{firstpage}

\begin{abstract}
The flux and the spectral index in X-ray energy band from the knots of M87 jet as observed by 
{\it{Chandra}} indicate a possible synchrotron origin but cannot be explained by 
simple one zone models with continuous injection of non-thermal electrons. In 
this letter we propose a two-zone
model to explain the observed spectra of the knots of M87 jet.
We consider the synchrotron emission from a region with tangled magnetic field where
relativistic non-thermal electrons are continuously injected in from an associated acceleration 
region. The acceleration region is assumed to be compact zone possibly around a shock front. 
A power-law distribution of electron is injected into the acceleration region and are accelerated 
to a maximum energy determined by the acceleration time scale and the loss processes.
With the present model we are able to explain the overall broadband features of the 
knots of M87 jet. Also the present model predicts a change in spectral index at ultraviolet 
energies and future observations at these energies can be used to constrain the 
parameters involved in the model.

\end{abstract}


\section{Introduction}

M87 is a nearby giant elliptical galaxy (distance = $16$ Mpc [\cite{tonry}])possessing a 
one-sided jet with projected distance
$\approx$ 2 Kpc and bright in radio, optical and X-ray energies. The jet structure is very well
studied in radio (\cite{owen,bire,sparks}), IR (\cite{perl01,sparks}) and 
optical (\cite{meis,sparks}) bands. Prior to 
{\it{Chandra}}, M87 jet was not very well studied at X-ray energies due to the limited angular 
resolution of earlier X-ray telescopes, {\it{Einstein}} and ROSAT. However, {\it{Chandra}} due to its
better spatial resolution is able to resolve many fainter knots of M87 jet which are 
observed only in radio and optical bands earlier. Moreover, the 
position of these knots in X-ray energies are 
nearly coincident with their radio/optical counterparts(\cite{perl01,perl05}). 
Also M87 is the only radiogalaxy (other than CenA(\cite{sree})) 
which is detected in GeV-TeV energies(\cite{ahar03}).
Initially it was not well understood whether the TeV gamma-ray emission region 
is close to nucleus(\cite{georg})
or from the knot HST-1(\cite{staw06}). However the detection of M87 
by HESS confirmed the high energy emission from the region close to the 
nucleus(\cite{ahar06}). Recent VERITAS detection of VHE emission from M87 (\cite{wystan}) 
again suggests the emission may not be from 
HST-1\footnote{http://polywww.in2p3.fr/actualites/congres/blazars2008}. 
Considering the fact M87 jet is misaligned to the 
observer, explaining this TeV gamma-ray emission required a modified model 
other than the one used to explain blazar emission. \cite{nero} explained 
TeV gamma-ray emission due to radiative cooling of electrons accelerated by strong 
rotation induced electric fields in the vacuum gaps in black hole magnetospheres. 
\cite{lenain} proposed a 
multi-blob model with several plasma blobs moving in the large opening angle of the jet 
formation zone. TeV gamma-ray emission is explained as Doppler boosted synchrotron self Compton 
radiation by the blobs moving close to the line of sight.

The radio-to-optical emission from the knots of M87 jet are quite well accepted as synchrotron 
emission due to cooling of relativistic non-thermal electrons by the 
magnetic field therein(\cite{perl01}). 
The flux and the spectral indices at X-ray energies indicates a possible continuation of synchrotron 
emission of the radio-to-optical spectra with the change in the spectral index 
beyond optical energies(\cite{marshall}, \cite{staw05}). 

Simple theoretical models viz. continuous injection model (hereafter CI)
(\cite{karda62,ginz68,heav87,meisen89}) 
and one time injection model (\cite{jaffe,karda62,pach70})
were unable to explain the observed X-ray flux and/or the spectral index. 
The CI model considers the synchrotron emission from non-thermal electrons 
injected continuously into a region with tangled magnetic field. But the 
X-ray flux predicted by this model is more than the 
observed flux(\cite{perl05}). In one time injection model, 
a burst of electrons are injected at $t=0$  and 
allowed to evolve with (\cite{jaffe}) or without pitch angle 
scattering (\cite{karda62,pach70}). The model with pitch angle scattering
under predicts the X-ray flux and the one without pitch angle scattering fail to predict the
observed X-ray spectral index (\cite{perl05}). 
However CI and one time injection models are able to reproduce the broadband spectra of 
the knots of various AGN detected by{\it{Chandra}} (\cite{saha03,samb}). 

In this letter we propose a two zone model to explain the non-thermal emission from the 
knots of M87 jet. We consider a situation where a power-law distribution of electrons 
are accelerated by an shock. The initial particle spectrum which is accelerated by the
shock can be an outcome of an
earlier acceleration process. The particles accelerated by the shock
cool via synchrotron radiation in a homogeneous magnetic field behind it. 
The present model is similar to a situation where particles accelerated by an external
shock are advected downstream to be accelerated further by an internal shock.
Two zone models were in fact used by various authors to explain the spectral and the temporal 
behaviour of blazars (\cite{kirk,subir,li}). 
In the next section, the formulation of the present model and the parameters involved are discussed. 
In \S 3 we discuss the results of the fitting using the present model and compare the present 
model with the other existing models. 

\section[]{The Two Zone Model}

We consider the acceleration of a power-law distribution of particles (which may be a relic 
of past acceleration process) at a shock front and cool via synchrotron radiation in a 
homogeneous magnetic field. We treat the present scenario as a two zones: one around
the shock front where the particles are accelerated (AR) and downstream region where they
loose most of their energy through synchrotron process (CR). This model is then used to 
explain the radio-optical-X-radio spectra of the knots in M87 jet.
We assume the CR to be a spherical blob of radius $R$ with tangled magnetic 
field $B_{{CR}}$ and AR is assumed to be a very thin region with magnetic field $B_{AR}$. 
Power law distribution of electrons are continuously injected into AR and are accelerated 
by the shock characterized by an acceleration timescale $t_{acc}$. Particles are 
then accelerated at an rate $1/t_{acc}$ to a maximum energy determined by the loss processes.
AR is assumed to be compact and the emission from CR mainly contributes the overall photon spectrum. 

The kinetic equation governing the evolution of electrons in AR is given by

\begin{equation}
\label{eq:arkin}
{\partial n(\gamma,t) \over \partial t} = 
{\partial  \over \partial \gamma} \left[\left(\beta_{AR} \gamma^2 
- \frac{\gamma}{t_{acc}}\right)n(\gamma,t)\right] 
- \frac{n(\gamma,t)}{t_{esc}} + Q (\gamma) 
\end{equation}
Where 
\begin{equation}
Q (\gamma) d\gamma = q_o \gamma^{-p} d\gamma \;\;\; \hbox  {for}\;\;\; \gamma_{min}<\gamma<\gamma_b
\end{equation}
Here $\gamma$ is the Lorentz factor of the electron, $t_{esc}$ is the escape timescale and 
$\beta_{AR} = \frac{\sigma_T}{6\pi m c}B_{AR}^2$.

Equation (\ref{eq:arkin}) can be solved analytically using Green's function 
(\cite{atoyan}) and the electron distribution for an energy-independent $t_{acc}$ and 
$t_{esc}$ at time $t$ is given by

\begin{align}
\label{eq:arnum}
n(\gamma,t) = t_{acc} \gamma^{-(\alpha+1)} \left(1-\frac{\gamma}{\gamma_{max}}\right)^{\alpha-1}
\nonumber \\
\int\limits_{x_o}^{\gamma} Q(x) \left[\frac{1}{x}-\frac{1}{\gamma_{max}}\right]^{-\alpha} dx
\end{align}
Where $\alpha = t_{acc}/t_{esc}$ and the lower limit of integration $x_o$ is given by

\begin{equation}
\label{eq:arlimit}
x_o=\left[\frac{1}{\gamma_{max}}+\left(\frac{1}{\gamma}-\frac{1}{\gamma_{max}}\right)exp(t/t_{acc})\right]^{-1}
\end{equation}
$\gamma_{max}=1/(\beta_{AR}t_{acc})>\gamma_b$ is the maximum Lorentz factor an electron can 
attain in AR. For $t>>t_{acc}$ equation (\ref{eq:arlimit}) can be approximated 
to be $\gamma_{min}$ 
as the injection term in equation (\ref{eq:arnum}) vanishes for $x<\gamma_{min}$. 

The evolution of the electrons in CR is governed by the equation
\begin{equation}
\label{eq:crkin}
{\partial N(\gamma,t) \over \partial t} = 
{\partial  \over \partial \gamma} \left[\beta_{CR} \gamma^2 N(\gamma,t)\right] 
+ Q_{AR}(\gamma) 
\end{equation}
Here the last term is the injection from AR $Q_{AR}(\gamma) = n(\gamma)/t_{esc}$ and for $t>>t_{acc}$
\begin{align}
Q_{AR}(\gamma) \approx q_o\alpha \gamma^{-(\alpha+1)}\left(1-\frac{\gamma}{\gamma_{max}}\right)^{\alpha-1} 
\nonumber\\
\int\limits_{\gamma_{min}}^{MIN(\gamma,\gamma_b)} x^{-p} \left[\frac{1}{x}-\frac{1}{\gamma_{max}}\right]^{-\alpha} dx
\end{align}
and $\beta_{CR} = \frac{\sigma_T}{6\pi m c}B_{CR}^2$.

The distribution of electron at time $t$ in CR from equation (\ref{eq:crkin}) is given by

\begin{equation}
N(\gamma,t) = \frac{1}{\beta_{CR}\gamma^{2}}\int\limits_{\gamma}^{\Gamma_o} Q_{AR}(x) dx
\end{equation}
Where $\Gamma_o = \gamma/(1-\gamma\beta_{CR}t)$.

From equation (\ref{eq:arnum}), it can be shown that the injection into CR 
(for $\alpha+1>p$) is a broken power-law
with index $-p$ for $\gamma<\gamma_b$ and $-(\alpha+1)$ for $\gamma>\gamma_b$. The synchrotron
losses in CR introduces an additional break $\gamma_c$ in the electron spectrum depending upon
the age of CR ($t_{obs}$) and the $B_{CR}$. 

\begin{equation}
\gamma_c = \frac{1}{\beta_{CR}t_{obs}}
\end{equation}

The electron spectrum in CR at $t_{obs}$ can 
then have two different spectral shapes depending on the location of $\gamma_c$ with respect
to $\gamma_b$.

\noindent (i) $\gamma_c > \gamma_b$: The final spectrum will have two breaks with indices 
\begin{equation}
\label{eq:crspec1}
N(\gamma,t_{obs}) \propto \left\{
\begin{array}{ll}
\gamma^{-p},&\mbox {~$\gamma_{min}<\gamma<\gamma_b$~} \\
\gamma^{-(\alpha+1)},&\mbox {~$\gamma_b<\gamma<\gamma_c$~} \\
\gamma^{-(\alpha+2)},&\mbox {~$\gamma_c<\gamma<\gamma_{max}$~} 
\end{array}
\right.
\end{equation}
\noindent (ii) $\gamma_c < \gamma_b$: In this case the indices are 
\begin{equation}
\label{eq:crspec2}
N(\gamma,t_{obs}) \propto \left\{
\begin{array}{ll}
\gamma^{-p},&\mbox {~$\gamma_{min}<\gamma<\gamma_c$~} \\
\gamma^{-(p+1)},&\mbox {~$\gamma_c<\gamma<\gamma_b$~} \\
\gamma^{-(\alpha+2)},&\mbox {~$\gamma_c<\gamma<\gamma_{max}$~} 
\end{array}
\right.
\end{equation}

The synchrotron emissivity $\epsilon_\nu$ from the resultant electron spectrum is 
then calculated by convoluting $N(\gamma,t)$ with single particle 
emissivity averaged over an isotropic distribution of pitch angles $P(\nu,\gamma)$ 
\begin{equation}
\epsilon_\nu = \frac{1}{4\pi}\int\limits_{1}^{\infty}N(\gamma,t)P(\nu,\gamma) d\gamma
\end{equation}

The predicted spectrum by the above model depends on nine parameters, which are $q_o$, $\alpha$,
$\gamma_{min}$, $\gamma_b$, $\gamma_{max}$, $p$, $B_{CR}$, $R$ and $t_{obs}$.
Here $\alpha$ and $p$ are estimated from the radio-to-optical and optical-to-X-ray spectral indices,
$q_o$ and $B_{CR}$ are constrained using the observed luminosity and equipartition magnetic field.
For $R$ we assume the physical sizes measured in radio (\cite{hardee}). Age of the knot $t_{obs}$
is chosen to introduce a break in the observed spectrum at optical band and $\gamma_b$ is fitted to
reproduce the observed X-ray flux. $\gamma_{min}$ and $\gamma_{max}$ are used as free parameters and
are fixed at $5$ and $10^8$.

\begin{figure}
\includegraphics[height=0.5\textheight, width=0.5\textwidth] {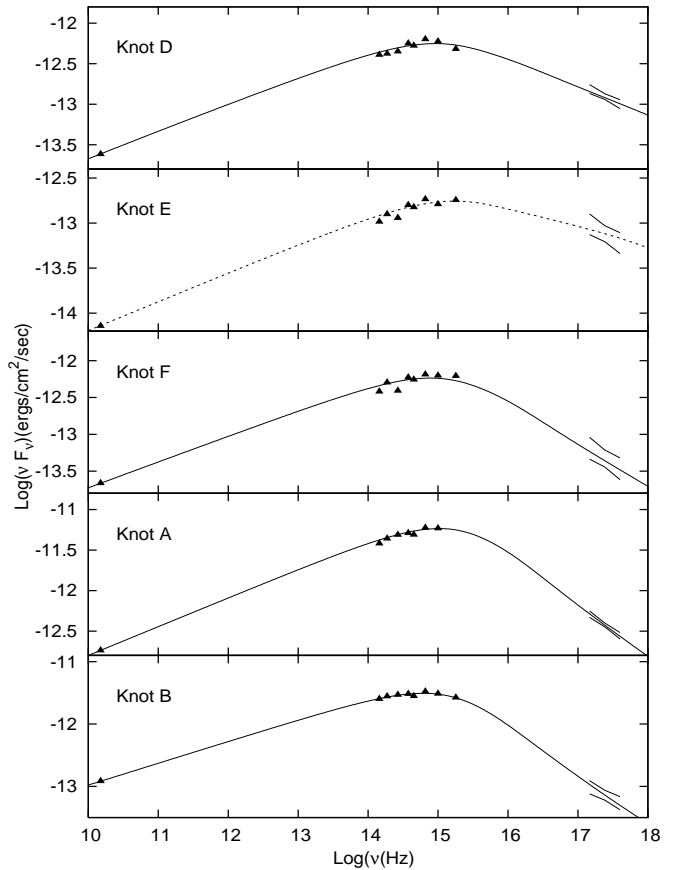}
\caption{ The spectral fit of the present model with the observed fluxes in radio(Perlman et al.(2001)), 
optical(Perlman et al.(2001), Waters $\&$ Zepf(2005)) and X-ray((Perlman $\&$ Wilson(2005)). 
The observed spectra of Knot E is fitted with 
CI model (dashed line). Errors in radio and optical are too small to be seen and hence not
shown in the figure.}\label{fig}  
\end{figure}

\section{Results and Discussion} 

We applied the above model to explain the knots D, F, A and B 
of M87 jet and the results of the fitting are shown in Figure~\ref{fig}
and the parameters used for the fit are given in Table 1. 
The spectrum of knot E can be explained by simple CI model and the parameters
we quote corresponds to CI model. 
We did not model knot C
due to significant differences in X-ray-optical properties(\cite{perl05}).

For all the fits shown in Figure~\ref{fig}, $\gamma_c < \gamma_b$. However, one can 
fit the spectrum with  $\gamma_c > \gamma_b$ with proper
choices of the parameters $\alpha$, $\gamma_b$ and $\gamma_c$.
This degeneracy arises due to the unavailability of UV spectral index since
the present model predicts the corresponding particle spectral index 
as $-(\alpha+1)$ or $-(p+1)$ (equation (\ref{eq:crspec1}) and (\ref{eq:crspec2}))
depending upon the above two conditions. 
Future observations at these photon energies may help in validating the present model 
and also will remove this degeneracy.
Also, to obtain a precise values for
$p$ and $\alpha$, spectral indices at radio-optical and X-ray energies should be
known accurately.  
In general $t_{acc}$ and $t_{esc}$ can be energy dependent and in such a situation
the solution (equation (\ref{eq:arnum})) may differ from its form and 
the index beyond $\gamma_b$
may not be the one discussed above.

Using the present model we estimated the flux at GeV-TeV energies due to inverse
Compton scattering and are given in Table 1. For the inverse Compton process the 
target photon can be 
either synchrotron photon(SSC) or cosmic microwave background radiation(IC/CMBR) or both.
From the observed superluminal velocities of $2c$ to $6c$ and the viewing angle of
$10^o$ to $20^o$ (\cite{biretta99}) we found the synchrotron photon energy density
dominates over the Doppler boosted cosmic microwave background radiation. Hence
the flux quoted in Table 1 are due to SSC process and are below the sensitivity 
of the present gamma-ray telescopes {\it{MAGIC}}(\cite{pratik}), {\it{HESS}}
and the upcoming experiment {\it{GLAST}}
{\footnote{http://www-glast.slac.stanford.edu/}}. Hence discrimination of the models 
based on GeV-TeV observations as discussed in \cite{georg06} for $3C 273$ cannot be
done for M87 jets.

A possible scenario 
of the present model is where AR is a region around an internal shock following an external shock. 
The electron injection
into AR can be the one which are already accelerated by an external shock and are advected
downstream to be accelerated further by the internal shock (\cite{tammi,pope}). 
An internal shock description
for the knots in AGN jets has already been discussed in literature (\cite{rees,saha05}).
Alternatively, reacceleration of power law electron distribution by turbulence at
boundary shear layers can also be another possible scenario (\cite{young,staw03}).
Inclusion of these scenario in its exact form into the present model 
will make it more complex and is beyond the scope of the present work. 

\cite{perl05} proposed a modified CI model 
where the volume within which particle acceleration occurs is energy-dependent. This
is expressed in terms of a filling factor {\it{$f_{acc}$}} which is the ratio 
between the observed flux to the flux predicted by the simple CI model . They found
{\it{$f_{acc}$}} declining with increasing distance from the nucleus suggesting particle
acceleration taking place in larger fraction of the jet volume in the inner jet than 
the outer jet. The energy dependence of {\it{$f_{acc}$}} also indicates that particle
acceleration regions occupy a smaller fraction of jet volume at higher energies. 
Even though the model is phenomenological, it indicates that the process of 
high energy emission from the knots are as complicated as their physical region. 
However the mechanism responsible for the filling factor is not explained.

\cite{staw06} explained the knot HST-1 of M87 jet as a region when the reconfinement
shock reaches the jet axis. They considered at the initial stage of M87 jet, the 
particles expand freely decreasing the pressure rapidly than the ambient gas pressure.
This will develop (in case of M87) a reconfinement shock which reaches the jet axis at 
a location which coincide with that of the knot HST-1. They postulate this location as the
beginning of HST-1 and while its outer parts are identified as stationary reflected 
shock formed when the reconfinement shock reaches the jet axis. Also they evaluated the 
ambient radiation field along the jet axis and estimated the TeV gamma-ray emission from 
HST-1 initiated by an outburst experienced at the core. Since the reconfinement shock
requires an initial free expansion, the knots downstream HST-1 cannot be explained by
this model.  

\cite{fleishman} explained the flattening of non-thermal spectra in the ultraviolet and 
X-ray bands observed from the 
knots of M87 and 3C273 jets through diffusive synchrotron radiation(DSR) in 
random small-scale magnetic fields. Whereas the synchrotron spectrum from regular 
large-scale magnetic field dominates the spectra at low energy band. The DSR spectrum
at high energy is $\propto \omega^{-\nu}$ where $\omega$ is the observed photon frequency 
and $\nu$ is the spectral index of the random magnetic field assumed to be a power-law.
\cite{honda} proposed a filamentary jet model to explain the observed X-ray spectral
index.
In their model, the jet comprises magnetic filaments of transverse size $\lambda$ 
and particles trapped in this filaments are accelerated
by diffusive shock acceleration. 
The acceleration of the 
electrons bound to a large filament
are controlled by the radiative losses before escape from the filament. Whereas the electrons
trapped in smaller filaments escape via energization. A critical scale $\lambda_c$
discriminates between the large and small scale filaments. They considered  a 
situation where the magnetic field is larger for filaments with larger size and found the
electron energy peaks when trapped in the filament of size $\lambda_c$. The X-ray 
spectrum is explained by the synchrotron radiation of the electrons accelerated in 
the filaments of size $\lambda>\lambda_c$. However, synchrotron radiation from large-scale
magnetic field itself can reproduce the observed X-ray spectrum (present model) involving
less number of parameters and/or assumptions(\S 2).

Recently \cite{liu} proposed a two zone model to explain 
the observed spectra of the knots of M87 jet. In their model electrons are
accelerated to relativistic energies in acceleration region (AR) and loose
most of their energies in cooling region (CR) through synchrotron
process. They considered AR and CR are spatially separated and introduced 
a break in the particle spectrum injected in CR through the advection of 
particles from AR to CR. This along with the cooling break in CR produce a double
broken power-law with indices $-p$, $-(p+1)$ and $-(p+2)$ which is then used to fit the 
observed spectra. However the present model assumes AR and CR are co-spatial supporting a more
physical scenario where electrons accelerated by the shock, cools in its vicinity.

 \begin{table*}
  \centering
  \begin{minipage}{160mm}
   \caption{Model Parameters and Knot Properties}
   \begin{tabular}{cccccccccccc}
   \hline
    Knot & $q_o$ & $\alpha$ & $p$ & $\gamma_b$ & $B_{CR}$ & $t_{obs}$ & $\gamma_c$ & $R$ & 
$F_{1GeV}$ & $F_{50GeV}$ & $F_{1TeV}$\\ 
\vspace{0.0mm}
 & & & & & & & & & (photons/$cm^2$/s) & (photons/$cm^2$/s) & (photons/$cm^2$/s) \\
 \hline
D & 7.5 & 1.75& 2.35& 1.8 & 9.3 & 1.6 & 5.7 & 12 & $1.8\times10^{-12}$ & $3.7\times10^{-14}$ & $2.1\times10^{-16}$ \\
E & 4.7 & ... & 2.36& ... & 5.9 & 2.5 & 9.0 & 17 & $1.7\times10^{-13}$ & $4.1\times10^{-15}$ & $4.6\times10^{-17}$ \\
F & 0.6 & 2.3 & 2.3 & 2.1 & 4.5 & 4.7 & 8.2 & 29 & $1.1\times10^{-12}$ & $2.6\times10^{-14}$ & $1.0\times10^{-16}$ \\
A & 1.0 & 2.45& 2.29& 2.0 & 4.7 & 3.5 & 10.1& 55 & $2.3\times10^{-11}$ & $5.9\times10^{-13}$ & $2.1\times10^{-15}$ \\
B & 0.9 & 2.75& 2.3 & 1.4 & 4.7 & 3.9 & 9.0 & 50 & $1.0\times10^{-11}$ & $2.3\times10^{-13}$ & $3.8\times10^{-16}$ \\

\hline
 \end{tabular}
\\
Columns:- 1: Knot name. 2: Normalisation (in units of $10^{-12}$)
of the power-law injected into AR(for knot E it is the normalisation of the 
power-law injected into CR (see text)). 3: Ratio between the 
acceleration timescale and escape timescale in AR. 4: Index of the power-law spectrum
injected into AR. 5: Maximum energy of the electron Lorentz factor injected into AR
(in units of $10^{6}$). 6: CR magnetic field (in units of $10^{-4}$ G). 
7: Time of observation (in units of $10^{9}$ sec). 8: Break Lorentz factor of the
electrons due to synchrotron cooling in CR (in units of $10^{5}$). 9: Size of
CR (in parsec) measured in radio (\cite{hardee}). 10: Flux at 1GeV.
11: Flux at 50GeV. 12: Flux at 1TeV.
  
For all cases, the minimum Lorentz factor($\gamma_{min}$)injected into AR is $5$ and 
maximum Lorentz factor($\gamma_{max}$) attained in AR is $10^8$. 

\end{minipage}
 \end{table*}

\section{Conclusion}

The observed radio-optical-X-ray spectra from the knots in the jets of the FRI radio galaxy 
M87 are explained within
the framework of two zone model. We considered a power-law electron distribution which 
are further accelerated in an acceleration region  
and are injected into a cooling region where they lose their
energy through synchrotron radiation. In its simplest form, the model does not consider
any specific acceleration process but assumes an energy independent acceleration timescale.
Future observations of M87 knots in UV-to-X-ray photon energies will confirm the present
model and constrain the parameters involved.

We explored the possibility of the present model to reproduce the X-ray flux of 
other FRI galaxies (detected by {\it{Chandra}}) which are observed to have 
lower radio luminosity and relatively 
smaller jets when compared with FRII galaxies. The X-ray emission from FRI jet is quite
well accepted to be of synchrotron origin whereas for FRII and quasars it may be due to IC/CMBR.
However the latter is still under debate (see \cite{harris} for a review about the 
X-ray emission from extragalactic jets). The X-ray emission from the knots and/or the jets 
of the FRI galaxies viz. 3C 66B(\cite{hardcastle}), 3C 346(\cite{worall}), 
CenA (\cite{hardcastle06}) and 3C 296 (\cite{hardcastle05} listed in the online 
catalog of extragalactic X-ray jets 
XJET \footnote{http://hea-www.harvard.edu/XJET/}, which are not explained by synchrotron 
emission from simple one zone models, can be reproduced by the present model.

The author thanks S. Bhattacharyya, N. Bhatt and M. Choudhury for the useful discussions
and suggestions. The author is grateful to referee  E. Perlman for useful comments 
and suggestions. This work has made use of the XJET website.

\label{lastpage}

\end{document}